\documentclass[aps,prb,showpacs,showkeys]{revtex4-1}
\usepackage[T2A,OT1]{fontenc}
\usepackage[cp1251]{inputenc}
\usepackage[russian,ukrainian,english]{babel}
\usepackage{graphicx} 
\usepackage{latexsym}
\usepackage{amsfonts}
\usepackage{xcolor}[2007/01/21]
\usepackage[dvipdfm,bookmarks=true,bookmarksnumbered,bookmarksopen]{hyperref}
\hypersetup{pdfpagemode=FullScreen,colorlinks=true,
citecolor=green,unicode,bookmarksopenlevel=3,
pdftoolbar=true,pdfauthor={Helen Gomonay},pdftitle={Shape effect}}
\usepackage{amsbsy}         
\begin{document}

\selectlanguage{english} \sloppy
\title{Shape-induced anisotropy in antiferromagnetic nanoparticles}
\author{H. Gomonay, S. Kondovych and V. Loktev}
\affiliation{National Technical University of Ukraine
``KPI'', ave Peremogy, 37, 03056 Kyiv, Ukraine}

\begin{abstract}
High fraction of the surface atoms considerably enhances the
influence of size and shape on the magnetic and electronic
properties of nanoparticles. Shape effects in ferromagnetic
nanoparticles are well understood and allow to set and control the
parameters of a sample that affect its magnetic anisotropy during
production. In the present paper we study the shape effects in the
other widely used magnetic materials -- antiferromagnets, -- which
possess vanishingly small or zero macroscopic magnetization. We
take into account the difference between the surface and bulk
magnetic anisotropy of a nanoparticle and show that the effective
magnetic anisotropy depends on the particle shape and
crystallographic orientation of its faces. Corresponding
shape-induced contribution to the magnetic anisotropy energy is
proportional to the particle volume, depends on magnetostriction,
and can cause formation of equilibrium domain structure.
Crystallographic orientation of the nanoparticle surface
determines the type of domain structure. The proposed model allows
to predict the magnetic properties of antiferromagnetic
nanoparticles depending on their shape and treatment.
\end{abstract} \maketitle

\section{Introduction}
Magnetic nanoparticles (NP) are widely used as constitutive
elements for the information technology (e.g. memory cells, spin
valves, magnetic field controllers etc.). To drive and control the
magnetic state of a particle and values of critical fields and
currents, we can use not only internal properties of magnetic
material, but also shape and size of the sample. As for
ferromagnetic (FM) particles, shape effects allow to tailor the
effective magnetic anisotropy and  critical field values during
production.

On the other hand, nowadays technologies use antiferromagnetic
(AFM) nanoparticles along with (or sometimes instead of) FM ones.
Experiments with AFM particles show that the reduction of size to
tens of nanometres leads to noticeable changes of properties
compared to the bulk samples: increase of lattice parameters in
the magnetically ordered phase \cite{Petrika:2010, Petrik:2011(2),
NegThermoEx}; increase of the magnetic anisotropy
\cite{Petrik:ZAAC201008027}; pronounced decrease of AFMR frequency
\cite{Bahl2006398}. Some of the finite size effects could be
caused by the shape and faces orientations of nanoparticle. For example, according to the N\'{e}el predictions
\cite{Neel:1949, Neel:1962}, small AFM particles exhibit
uncompensated magnetic moment with the size- and shape-dependent
value \cite{Kodama:PhysRevLett.77.394,
Morup:springerlink:10.1007/978-94-010-0045-1_33}. Recent
experiments with rather large (100-500 nm size) AFM particles
\cite{Scholl:d10.1021/nl300361e, Scholl:PhysRevB.84.220410,
Scholl:doi:10.1021/nl1025908} discovered the shape effects similar
to the shape-induced phenomena in FM materials: \emph{i})
switching of AFM vector from crystallographic to particle
easy-axis with an increase of aspect ratio; \emph{ii})
correlation between the type of domain structure and such parameters as aspect ratio
of the sample and orientation of faces.

However, the mechanism of the finite-size and shape effects in AFM
nanoparticles is still an open issue.

Shape effects in AFM particles could, in principle, originate from
a weak ferromagnetic moment \cite{Garanin:2003PhRvL..90f5504G}
thus reducing the difference between AFM and FM systems to
quantitative one. On the other hand, certain dynamic and
equilibrium properties of AFM, like peculiarities of the magnon spectra  or coupling to the external magnetic field,
could not be reduced to FM ones.

Understanding the mechanisms of the shape effects specific to AFM
ordered systems is crucial for optimizing and finetuning the
properties of AFM-based devices and clarifying the  fundamental
questions whether the shape effects reside in AFM with vanishingly
small macroscopic magnetization,  and which of peculiar AFM
properties might depend on the particle shape. For this purpose we
investigate the finite-size and shape effects in AFM particles,
regardless of their macroscopic magnetization, combining two
previously shown statements: \emph{i}) the shape effects in AFM
materials may originate from the long-range fields of
``magnetoelastic'' charges due to spontaneous magnetostriction
below the N\'eel temperature (so called destressing fields)
\cite{gomo:PhysRevB.75.174439}; \emph{ii}) ``magnetoelastic''
charges may arise from the surface magnetic anisotropy
\cite{Gomonay:cmjp2010}. We consider the particles with the
characteristic size below the several critical lengths of
monodomainization (which, for convenience, are referred to as
``nanoparticles'').

The basic idea is to consider \emph{a priori} the surface and bulk
properties as different: to distinguish the constants of surface
and bulk magnetic anisotropy and, as a consequence, equilibrium
orientation of AFM vectors at the surface and in the bulk. We show
that due to the long-range nature of elastic forces, the surface
anisotropy contributes to the magnetic energy of the sample. This
contribution is proportional to the particle volume, depends on
the aspect ratio and crystallographic orientation of the sample
faces, and affects equilibrium (single- and multi-domain) state
of AFM nanoparticle.

The proposed approach requires consistent description of the
magnetic and elastic subsystems of AFM particles and thus differs
from the well-known formalism for the FM
\cite{Kittel:RevModPhys_21_541}. M.I. Kaganov \cite{Kaganov:1980R}
has already pointed out the role of the surface magnetic
anisotropy effects on spin-flip transitions in magnetic materials,
considering the magnetic moment at the surface as an additional
parameter; however, his approach eliminates magnetoelastic and
shape effects, while in our work we assume noticeable influence of
these effects on the properties of the nanosized AFM particle.

\section{Model}
To describe the equilibrium magnetic state of a NP we need to
introduce at least three additional (in comparison with bulk
samples) parameters that characterize: \emph{i}) shape, \emph{ii})
size, and \emph {iii}) orientation of sample faces. 

\begin{figure}[htbp]
\begin{centering}
\includegraphics[width=0.5\textwidth]{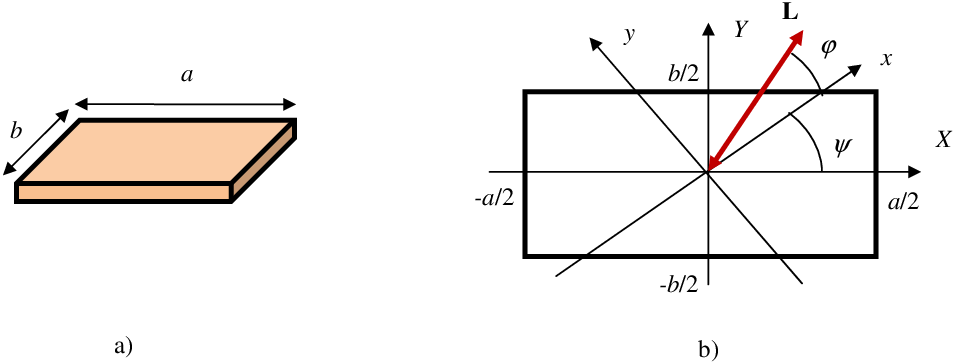}
  \caption{(Color online) Sample
  (a) and orientation of the N\'eel vector $\mathbf{L}$ (b) with respect to crystal axes ($x,y$)
  and sample edges ($X,Y$).
 \label{fig_sample}}
 \end{centering}
\end{figure}

We consider a thin flat rectangular particle (thickness $h\ll
b<a$, Fig.~\ref{fig_sample}), typical for experimental studies
(see, e.g., \cite{Scholl:doi:10.1021/nl1025908}). The thickness
$h$ of a particle is, however, large enough to ensure an AFM
ordering (i.e., significantly larger than the magnetic correlation
length).

The sample surface (see Fig.~\ref{fig_sample} b) consists of four
faces with the normal vectors $\mathbf{n}_j=(\cos\psi_j,
\sin\psi_j)$, $j = 1,\ldots, 4$ ($x, y$ are parallel to
crystallographic axes). We disregard the upper and lower faces ($z=Z =
const$) as they do not contribute to the effects discussed below.
Equations that define the surface are ($Z \in [0, h]$, $X, Y$ are
parallel to the particle edges):
\begin{equation}\label{eq_surface}
    \begin{array}{llll}
     X=a/2,&Y\in[-b/2,b/2],&\mathbf{n}_1=(1, 0),& \psi_1=\psi, \\
    Y=b/2,&X\in[-a/2,a/2],&\mathbf{n}_2=(0, 1),& \psi_2=\psi+\pi/2,\\
   X=-a/2,&Y\in[-b/2,b/2],&\mathbf{n}_3=(-1, 0),&  \psi_3=\psi+\pi, \\
     Y=-b/2,&X\in[-a/2,a/2],&\mathbf{n}_4=(0, -1),& \psi_4=\psi+3\pi/2.\\
    \end{array}
\end{equation}

For such a model, the additional external (in thermodynamic sense)
parameters of the NP are: \emph{i}) aspect ratio $a/b$ (defines
the shape), \emph{ii}) width $b$ (defines characteristic size),
and \emph{iii}) angle $\psi$ (defines the orientation of the
surfaces).

We consider a typical collinear AFM with two equivalent
sublattices $ \mathbf{M} _1 $ and $ \mathbf{M} _2 $; the N\'eel
(or AFM) vector $\mathbf{L} = \mathbf{M} _1-\mathbf{M} _2$ plays a
role of the order parameter. Far below the critical point the
magnitude of the AFM vector is fixed (we assume $|\mathbf{L}|=1$).

\begin{figure}[htbp]
\begin{centering}
   \includegraphics[width=0.5\textwidth]{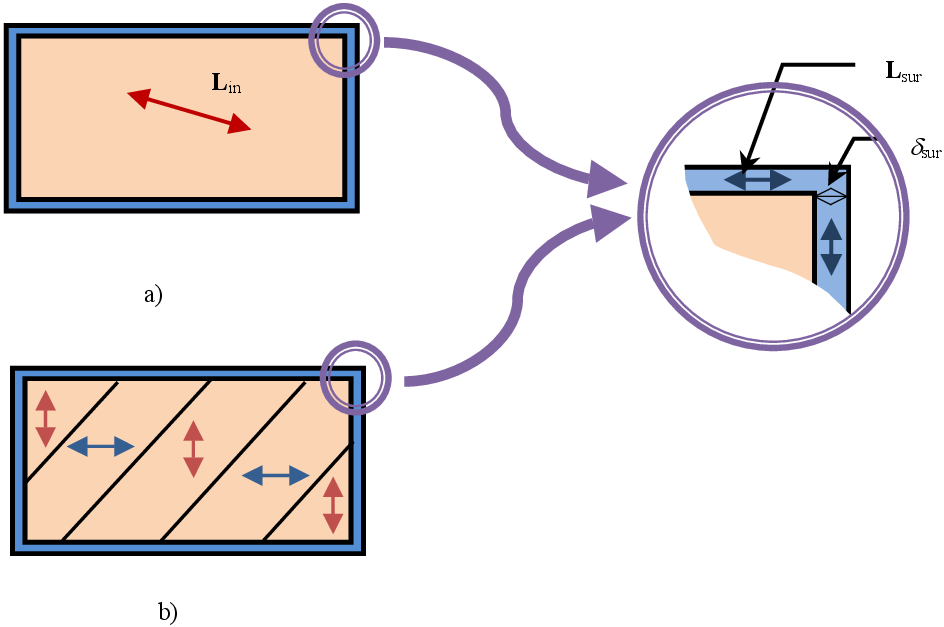}
  \caption{(Color online) Space distribution of AFM vector (arrows)
  in a single domain (a) and multidomain (b) states.
 \label{fig_domain_structure}}
 \end{centering}
\end{figure}

To obtain equilibrium distribution $\mathbf{L}(\mathbf{r})$
for the NP of given shape and size, we minimize the total energy 
$W$ which includes several terms of different nature. First of all, we may consider the surface as a separate magnetic phase
\cite{Perez:2008Nanot..19U5704P, Bhowmik:PhysRevB.69.054430,
Tobia:PhysRevB.78.104412}  with a small but finite thickness
$\delta_\mathrm{sur}$ (narrow peripheral region $S$ of thickness $\delta_\mathrm{sur}$
in Fig.~\ref{fig_domain_structure})  and thus distinguish the  bulk, $W_\mathrm{bulk}$, and the surface, $W_\mathrm{sur}$, contributions:
\begin{equation}\label{eq_free_energy_total}
W=W_\mathrm{bulk}+W_\mathrm{sur}.
\end{equation}

Then, we can also distinguish different contributions to the bulk energy, $W_\mathrm{bulk}$, the most important are those that describe the magnetic anisotropy, $w_{\mathrm{anis}}$, exchange, $w_\mathrm{exch}$, and magnetoelastic, $w_\mathrm{m-e}$ coupling. 
We assume that the bulk magnetic anisotropy corresponds to
tetragonal symmetry with two equivalent easy directions in the NP
plane ($x$ or $y$ in Fig.~\ref{fig_sample} b) and model respective
contribution to the energy density as:
\begin{equation}\label{eq_magnetic}
    w_{\mathrm{anis}}=\frac{1}{2}K_\| L_z^2-K_\perp
    (L_x^4+L_y^4),
\end{equation}
where $K_\| \gg K_\perp>0$ are the phenomenologic anisotropy
constants.

Exchange interactions (responsible for inhomogeneous distribution
of the N\'eel vector inside the sample) give rise to a gradient
term
\begin{equation}\label{eq_exchange}
w_\mathrm{exch}=\frac{1}{2}\alpha(\nabla \mathbf{L})^2,
\end{equation}
where $\alpha$ is a phenomenological constant. Competition between
the exchange coupling (\ref{eq_exchange}) and magnetic anisotropy
(\ref{eq_magnetic}) defines the characteristic size $\xi_{DW}$ of
the domain wall (DW): $\xi_{DW} = (1/2) \sqrt{\alpha/K_\perp}$.

Magnetoelastic coupling in AFM materials can be pronounced
pronounced  (compared with FM ones) due to the
presence of strong crystal field and, as a result, strong
spin-orbit coupling (like in oxides LaFeO$_3$ or NiO). In the
simplest case of the elastically isotropic material, the density
of magnetoelastic energy is:
\begin{equation}\label{eq_magnetoelastic}
    w_\mathrm{m-e}=\lambda_\mathrm{iso} \mathbf{L}^2 \mathrm{Tr}\hat u +2\lambda_\mathrm{anis} \left[(\mathbf{L}\otimes\mathbf{L}-\frac{1}{3}\hat I)(\hat u-\frac{1}{3} \hat I \mathrm{Tr}\hat u )\right],
\end{equation}
where $\hat u$ is the strain tensor, $\hat I$ is the identity
matrix, constants $\lambda_\mathrm{iso}$ and
$\lambda_\mathrm{anis}$ describe isotropic and shear
magnetostriction, respectively.

Final expression for the bulk energy is thus given by
\begin{equation}\label{eq_free_energy}
W_\mathrm{bulk}=\int_V(w_\mathrm{anis}+w_\mathrm{exch}+w_\mathrm{m-e}+w_\mathrm{elas})dV,
\end{equation}
where $w_\mathrm{elas}$ is the elastic energy density (see,
e.g.\cite{Landau:1987R}), $V=abh$ is the NP volume, all other
terms are defined above.

At last, let us focus on the magnetic surface energy $W_\mathrm{sur}$ which is of crucial
importance for our model and needs special discussion. Experiments with
the nanoscale AFM particles reveal significant difference between the
magnetic ordering and hence the magnetic properties of the surface
from those of the bulk. In
particular, depending on the material, treatment, and other
technological factors the NP surface may lack the long-range
magnetic structure (paramagnetic or spin glass
\cite{Tobia:PhysRevB.78.104412, Mishra:2008arXiv0806.1262M}), or
may have different type of ordering (e.g. multi- vs.
two-sublatteral in the bulk \cite{Kodama:PhysRevLett.79.1393}), or
different easy axis/axes. We consider the last case and assume, for the sake of simplicity, that
the easy magnetic axis at the surface is perpendicular to the normal
$\mathbf{n}$; then, expression for the magnetic surface energy
takes the form:
\begin{equation}\label{eq_surface_energy}
W_\mathrm{sur}= K_\mathrm{sur} \oint_S(\mathbf{L}\mathbf{n})^2 dS=K_\mathrm{sur}
\sum_{j=1}^4\int_{S_j}(\mathbf{L}\mathbf{n_j})^2dS,
\end{equation}
where $K_\mathrm{sur}>0$ is a phenomenological constant.
$W_\mathrm{sur}$ obviously depends on orientation of edges: angles
$\psi_j$, or, equivalently, vectors $\mathbf{n}_j$ (see (\ref{eq_surface})).

It is instructive to compare the specific surface magnetic anisotropy
$K_\mathrm{sur}/\delta_\mathrm{sur}$ with the magnetic anisotropy constant
$K_\perp$: they have the same order
of value ($K_\mathrm{sur}/\delta_\mathrm{sur} \propto K_\perp$),
if the broken exchange bonds play the main role in formation of
the surface properties; while in the case of dominating
dipole-dipole interactions $K_\mathrm{sur}/\delta_\mathrm{sur}$
can be much greater than $K_\perp$
\cite{Garanin:2003PhRvL..90f5504G}.

It should be stresses, that, in principle, all introduced phenomenological constants fall into two
categories: internal\footnote {~The values of the internal
constants in NPs could differ significantly (by several times or
even orders) from those for the bulk samples \cite{Petrika:2010,
NegThermoEx, Sun:0022-3727-42-12-122004}, where the role of the
surface is negligible.} (indexed ``$\mathrm{in}$'') and
superficial (indexed ``$\mathrm{sur}$''); interactions of both
types can contribute to the shape effects. However, in our model we
distinguish only between the magnetic constants
$K_\mathrm{sur}/\delta_\mathrm {sur}$ and $K_\perp$, taking this
difference as the most important.

The expression (\ref{eq_free_energy_total}) for the NP energy is the
functional over the field variables $\mathbf{L}(\mathbf{r})$ (the
AFM vector) and $\mathbf{u}(\mathbf{r})$ (the displacement
vector). We reduce the number of independent variables to three
assuming that: \emph{i}) vector $\mathbf{L}$ lies within the $xy$
plane and can be parametrised by a single angle $\varphi$ (see
Fig.~\ref {fig_sample} b) because of a strong
easy-plane anisotropy ($K_\| \gg K_\perp$); \emph{ii}) strain component $u_{zz} $ in a
rather thin plate ($h \ll a,b$) can be considered as homogeneous
and thus can be excluded from consideration (see
\cite{Landau:1987R}). The standard minimum conditions generate the
set of differential equations for one magnetic,
$\varphi(\mathbf{r})$, and two elastic, $u_x(\mathbf{r})$,
$u_y(\mathbf{r})$ variables in the bulk:
\begin{eqnarray}
&&  -\alpha \Delta \varphi+K_\perp\sin4\varphi+2\lambda_\mathrm{anis}
[-(u_{xx}-u_{yy})\sin2\varphi+2u_{xy}\cos2\varphi]= 0,
\label{eq_system_mag}\\
  &&\Delta u_x +\nu_{\mathrm{eff}}\nabla_x\mathrm{div}\mathbf{u}= -(\lambda_\mathrm{anis}/\mu)[\nabla_x\cos2\varphi+\nabla_y\sin 2\varphi], \label{eq_system_el1} \\
&&   \Delta u_y +\nu_{\mathrm{eff}}\nabla_y\mathrm{div}\mathbf{u} = -(\lambda_\mathrm{anis}/\mu)[\nabla_x\sin2\varphi-\nabla_y\cos 2\varphi].  \label{eq_system_el2}
\end{eqnarray}
Here, operators $\Delta$ and $\mathrm{div}$ are two-dimensional,
$\nu_{\mathrm{eff}}\equiv (1+\nu)/(1-\nu)$ is the effective
Poisson ratio (instead of 3-dimensional $\nu$
\cite{Landau:1987R}), $\mu$ is the shear modulus.

Equations for the AFM vector at the $j$-th face (variables
$\varphi^{(j)}_\mathrm{sur}$, see (\ref{eq_surface}))
\begin{equation}\label{eq_boundary_condition}
    -K_\mathrm{sur}\sin 2(\varphi^{(j)}_\mathrm{sur}+\psi_j)+\alpha (\mathbf{n}_j\nabla)\varphi^{(j)}_\mathrm{sur}=0
\end{equation}
\noindent could be considered as the boundary conditions. They
differ from the standard boundary conditions for AFMs (see, e.g.,
\cite{Gann:1980R, Marchenko:PhysRevLett.97.067204}) due to the
presence of the additional surface term with $K_\mathrm{sur}$.

In the limit $K_\mathrm{sur}\rightarrow 0$ the solutions of
equations (\ref{eq_system_el1}), (\ref{eq_system_el2}),
(\ref{eq_boundary_condition}) are well known: the AFM vector
$\mathbf{L}(\mathbf{r}) = const$ lies along one of the easy axes
($\varphi_{\mathrm{in}} = 0 $ or $ \pi / 2 $), the displacement
vector $\mathbf{u}(\mathbf{r})$ generates the homogeneous field of the
magnetically-induced strain:
\begin{equation}\label{eq_spontan-def}
    u^{(0)}_{xx}-u^{(0)}_{yy}=-\frac{\lambda_\mathrm{anis}}{\mu}\cos2\varphi_{\mathrm{in}},\quad u^{(0)}_{xy}=-\frac{\lambda_\mathrm{anis}}{2\mu}\sin2\varphi_{\mathrm{in}}.
\end{equation}
In the massive (infinite) samples the spontaneous striction
(\ref{eq_spontan-def}) causes magnetoelastic gap in the spin-wave
spectrum (in assumption of ``frozen'' lattice), but does not
affect the equilibrium orientation of the AFM vector (all the
magnetostrictive terms in (\ref{eq_system_mag}) cancel out,
eliminating the shape effect).

For the finite-size samples with nonzero surface anisotropy
($K_\mathrm{sur}\ne0$) the easy direction at least in some
near-surface regions unavoidably differs from that in the bulk and so,
the spatial distribution of the AFM vector shoould be non-uniform. As a
result, the sources of the displacement field -- the non-zero gradient terms, or ``\emph{magnetoelastic  charges}'' -- appear in the r.h.s of equations
(\ref{eq_system_el1}) and (\ref{eq_system_el2}). In the following section we discuss this issue in
more details.

\section{Shape-induced anisotropy}
The consistent theory of shape effects in AFMs should account for
the long-range elastic and magnetoelastic interactions and thus
should rest upon the complete set of equations
(\ref{eq_system_mag})-(\ref{eq_system_el2}). However, the
displacement field $\mathbf{u}(\mathbf{r})$ can be formally excluded from consideration once the Green tensor
$G_{jk}(\mathbf{r},\mathbf{r}^\prime)$ for equations
(\ref{eq_system_el1}) and (\ref{eq_system_el2}) is known (see Appendix
\ref{sec_Green}). In this case the spatial distribution of the AFM
vector $\mathbf{L}(\mathbf{r})$ should minimize the energy
functional
\begin{equation}\label{eq_free_energy_2}
W[\mathbf{L}(\mathbf{r})]=\int_V(w_\mathrm{mag}+w_\mathrm{exch})dV+W_\mathrm{sur}+W_{\mathrm{add}},
\end{equation}
which includes the additional term of magnetoelastic nature:
  \begin{eqnarray}\label{eq_nonlocal_energy}
W_{\mathrm{add}}&=&\frac{2\lambda^2_{\mathrm{anis}}}{\mu}\int_V\int_V\nabla_m\left[L_j(\mathbf{r})L_m(\mathbf{r})\right]G_{kj}(\mathbf{r},\mathbf{r}^\prime)\nabla^\prime_l\left[L_k(\mathbf{r}^\prime)L_l(\mathbf{r}^\prime)\right]dVdV^\prime\nonumber\\
&+&\frac{2\lambda^2_{\mathrm{anis}}}{\mu}\oint_S\oint_S
L_j(\mathbf{r}_{\mathrm{sur}})L_m(\mathbf{r}_{\mathrm{sur}})G_{kj}(\mathbf{r}_{\mathrm{sur}},\mathbf{r}^\prime_{\mathrm{sur}})L_k(\mathbf{r}^\prime_{\mathrm{sur}})L_l(\mathbf{r}^\prime_{\mathrm{sur}})dS_mdS^\prime_l.
\end{eqnarray}

Analysis of the Exp.~(\ref{eq_nonlocal_energy}) shows that any
inhomogeneous distribution $\mathbf{L}(\mathbf{r})$ gives nonzero,
generally positive contribution to energy $W_{\mathrm{add}}$. Due
to the ``Coulomb-like'' nature of the elastic forces
($G_{jk}(\mathbf{r},\mathbf{r}^\prime)\propto
1/|\mathbf{r}-\mathbf{r}^\prime|$) this contribution scales as
sample volume $V$. In addition, nonlocality of the
$W_{\mathrm{add}}$ term turns equations (\ref{eq_system_mag}) to
integro-differential ones and thus complicates the problem.

In the present paper we propose the simplified approach to solve
equations (\ref{eq_system_mag}) - (\ref{eq_system_el2}) using the
following peculiar features of antiferromagnets.

First, we consider the magnetostriction of AFMs as a secondary
order parameter which means that in the thermodynamic limit (in neglection of
boundary conditions) the homogeneous spontaneous strains
(\ref{eq_spontan-def}) preserve the symmetry of the magnetically
ordered state and orientation of the easy axis. In addition, though usually
the magnetoelastic energy is comparable (up to the order of value)
to the 4-th order magnetic anisotropy (i.e., to $K_\perp$
constant), it can be much less than the uniaxial magnetic
anisotropy. Thus, assuming strong uniaxial surface anisotropy
$K_\mathrm{sur}\gg K_\perp\delta_\mathrm{sur}$, we can disregard
the influence of magnetoelastic strains on equilibrium orientation
of AFM vector at the surface. However, this assumption does not
restrict the relation between $K_\mathrm{sur}$ and the
characteristic DW energy $\sigma_{DW}$, because
$\sigma_{DW}\propto \sqrt{K_\perp J}a_{\mathrm{lat}}\gg
K_\perp\delta_\mathrm{sur}$ (where $J\gg K_\perp$ characterizes
the exchange coupling, $a_{\mathrm{lat}}$ is the lattice constant,
and we used the following relations: $\alpha\propto
Ja^2_{\mathrm{lat}}$, $\delta_\mathrm{sur}\propto
a_{\mathrm{lat}}$).

Second, we propose the following hierarchy of characteristic
length scales: the width of magnetic inhomogeneuity is much less
than the sample size, $\xi_{DW}\ll a,b$, but much greater than
interatomic distance, $\xi_{DW}\gg a_{\mathrm{lat}}$ (due to
exchange enhancement); the width of elastic inhomogeneuity has
interatomic scale and thus is much less than $\xi_{DW}$. Note, that
the value of $\xi_{DW}$ in nanoparticles with the large fraction of the surface atoms can be much smaller than that for the
bulk samples due to variation of
magnetoelastic and exchange coupling (see, e.g.
\cite{Petrik:2011}). Thus, inequality $\xi_{DW}\ll a,b$ keeps true
in a wide range of the sample dimensions down to tens of nanometers (below this range applicability of the
continual model is questionable).

\begin{figure}[htbp]
\begin{centering}
 \includegraphics[width=0.5\textwidth]{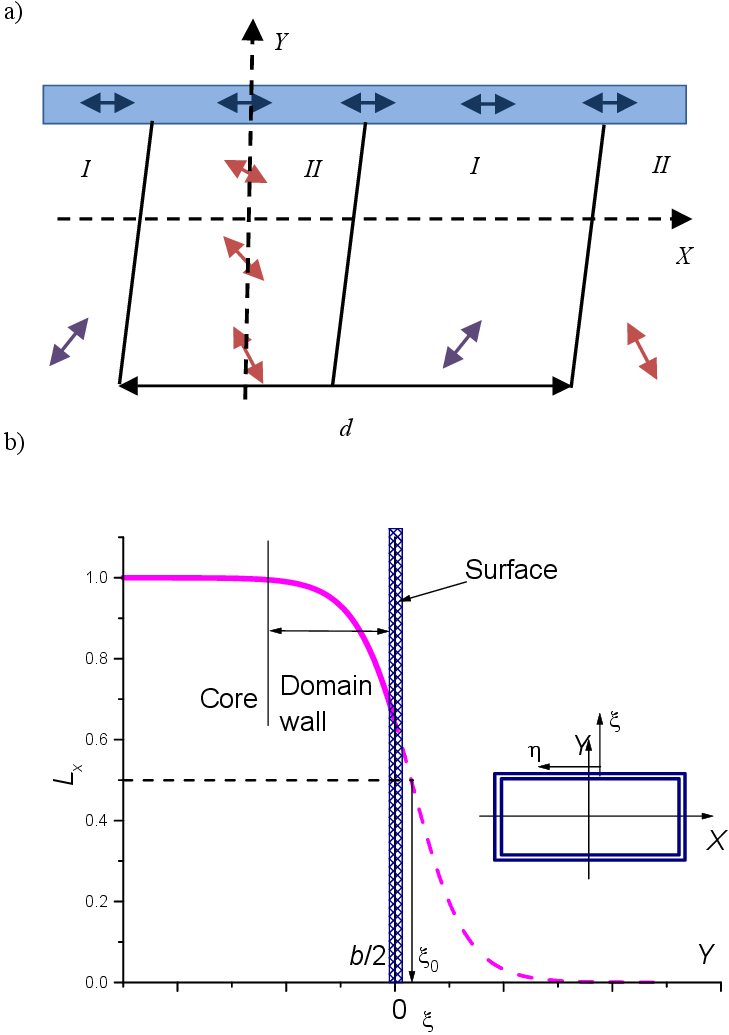}
  \caption{(Color online)  Distribution of the N\'eel vector in
  the vicinity of $Y = b/2$ face, multidomain state. (a) Periodic (period $d$) domain
structure, double arrows indicate orientation of AFM vectors
inside domains and in the near-surface region (shaded horizontal
stripe). (b) Space dependence  of $L_x (\xi)$ (solid line)
calculated from (\ref{eq_DW_profile}) provided that
$\varphi_{\mathrm{in}} = 0$. The horizontal line defines the
center $\xi_0$ of a virtual full domain wall (dotted line). Shaded
vertical bar indicates the position of surface region. Direction
of DW normal coincides with the axis $\xi$ of the local coordinate
system (inset). \label{fig_near_surface}}
 \end{centering}
\end{figure}

\renewcommand{\labelenumi}{\theenumi)}
Thus, within the above approximations, equilibrium orientation of the AFM vector at the surface results mainly from competition of the magnetic interactions: surface magnetic anisotropy and inhomogeneous exchange coupling, once the bulk vector $\mathbf{L}_{\mathrm{in}}$ is fixed. Orientation of $\mathbf{L}_{\mathrm{in}}$, in turn, is defined by interplay between the bulk magnetic anisotropy and magnetostrictive contribution  induced by spatial rotation of AFM vector in the thin ($
\propto \xi_{DW} $) near-surface region (see
Fig.~\ref{fig_near_surface}). So, the effective shape-induced
magnetic anisotropy and equilibrium distribution of AFM vector
could be determined self-consistently according to the following
procedure: \emph{i}) \label{item_one} to calculate
$\mathbf{L}_\mathrm{sur}$ starting from some (initially unknown)
``seed'' distribution of the AFM vector  $\mathbf{L}_\mathrm{in}$
in the NP bulk; \emph{ii}) \label{item_two} to substitute thus
defined seed distribution into equations for the displacement
vector and to determine corresponding field sources
(magnetoelastic charges); \emph{iii}) \label{item_three} to
calculate charge-induced average strains whose contribution
into free energy is proportional to the sample volume; \emph{iv})
\label{item_four} to define the effective magnetic anisotropy
which accounts for the average strains and calculate
$\mathbf{L}_\mathrm{in}$.

We note that the form of the seed distribution (and hence the free
variable of the structure) is different for a single- and a multi-domain
states. In the first case $\mathbf{L}_\mathrm{in}$ is homogeneous
within the bulk but can deflect from the
magnetic easy axis, so, $\varphi_{\mathrm{in}}$ is the
appropriate free variable. In the second case we assume, in
analogy with FM, that AFM vector within each of the domains
is fixed and parallel to one of two equivalent easy axes; then, free variable coincides with the fraction of type-I (or type-II) domains.

\subsection{Seed distribution and magnetoelastic charges}
In the simplest case of a single-domain state
(Fig.~\ref{fig_domain_structure} a), there are two homogeneous
regions: the ``shell'' (of the thickness $\delta_\mathrm{sur}$)
and the core. An equilibrium value  $\varphi_{\mathrm{in}}$ inside
the core is fixed, constant (as $\Delta \varphi_{\mathrm{in}}=0$),
but unknown (in some cases discussed below $\varphi_{\mathrm{in}}
= 0$ or $\pi/2$ that corresponds to one of the easy axes). We calculate the value
$\varphi^{(j)}_\mathrm{sur}$ at the surface from
Eq.~(\ref{eq_boundary_condition}) with account of the standard
expression for the domain wall profile:
\begin{equation}\label{eq_DW_profile}
    \sin 2\left(\varphi(\xi)-\varphi_{\mathrm{in}}\right)=2\xi_{DW}\frac{d\varphi}{d\xi}=\frac{1}{\cosh\left((\xi-\xi_0)/\xi_{DW}\right)}.
\end{equation}
Face normals generate the set of variables $\xi = (\pm X-a/2),
(\pm Y-b/2)$ of the local coordinate system (see inset in
Fig.~\ref{fig_near_surface} b). Position $\xi_0$ of the DW center
is calculated from the boundary conditions (see below). In
(\ref{eq_DW_profile}) we neglect the possible difference between
the DW width $\xi_{DW} = (1/2)\sqrt{\alpha/K_\perp}$ in the
near-surface region and in the core.

Substituting (\ref{eq_DW_profile}) in
(\ref{eq_boundary_condition}), we obtain the following equation
for $\varphi^{(j)}_\mathrm{sur}$:
\begin{equation}\label{eq_angle_boundary}
    \tan2\varphi^{(j)}_\mathrm{sur}=\frac{K_\mathrm{sur}\sin2\psi_j+\sigma_{DW}\sin2\varphi_{\mathrm{in}}}{\sigma_{DW}\cos2\varphi_{\mathrm{in}}-K_\mathrm{sur}\cos2\psi_j},
\end{equation}
where $\sigma_{DW} = \sqrt{\alpha K_\perp} = 2\xi_{DW} K_\perp$ is
the characteristic energy of the domain wall. The values
$\varphi^{(j)}_\mathrm{sur}$ at the opposite faces coincide:
$\varphi^{(1)}_\mathrm{sur} = \varphi^{(3)}_\mathrm{sur}$,
$\varphi^{(2)}_\mathrm{sur} = \varphi^{(4)}_\mathrm{sur}$ (see
(\ref{eq_surface})).

Analysis of Exp.~(\ref{eq_angle_boundary}) shows that the AFM
vector at the surface can be either
parallel to the edge: $\varphi^{(j)}_\mathrm{sur} = \psi_j$, as
shown in Fig.~\ref{fig_domain_structure}, (in the limit of large
surface anisotropy, $K_\mathrm{sur} \gg \sigma_{DW}$), or coincide
with the bulk AFM vector: $\varphi^{(j)}_\mathrm{sur}
=\varphi_{\mathrm{in}}$ (for the vanishing surface energy,
$K_\mathrm{sur} \ll \sigma_{DW}$). In the last case the surface influence and, correspondingly, shape
effects disappear.

Note that the surface DW is ``incomplete'': in general, DW center
is located outside the sample (see Fig.~\ref{fig_near_surface} b)
and its coordinate $\xi^{(j)}_0$ depends on the surface anisotropy
\begin{equation}\label{eq_center_DW}
    \xi^{(j)}_0=\xi_{DW}\sinh^{-1}\frac{K_\mathrm{sur}\sin2\psi_j+\sigma_{DW}\sin2\varphi_{\mathrm{in}}}{\sigma_{DW}\cos2\varphi_{\mathrm{in}}-K_\mathrm{sur}\cos2\psi_j}.
\end{equation}
In a single-domain nanoparticle the AFM vector rotates from
$\mathbf{L}_\mathrm{sur}$ to $\mathbf{L}_\mathrm{in}$ in a narrow,
almost zero-width ($ \le \xi_{DW}\ll a,b$) region and so, we can
model the spatial dependence of $\mathbf{L}(\mathbf{r})$ with a
step-like function. Within this approximation, r.h.s. of equations
(\ref{eq_system_el1}) and (\ref{eq_system_el2}) are nontrivial
only at the surface; this fact makes it possible to use a homogeneous form
of these equations for the bulk region of the NP:
\begin{equation}\label{eq_inner_elastic}
  \Delta \mathbf{u}
  +\nu_{\mathrm{eff}}\boldsymbol{\nabla}\mathrm{div}\mathbf{u}=0
\end{equation}
with the following boundary conditions for the displacement
vector:
\begin{equation}\label{eq_inner_elastic_boundary}
  (\mathbf{n}\cdot\boldsymbol{\nabla}) \mathbf{u}_\mathrm{sur}
  +\nu_{\mathrm{eff}}\mathbf{n}\mathrm{div}\mathbf{u}_\mathrm{sur}=\mathbf{n}\hat
  Q^{\mathrm{m-e}}.
\end{equation}
In (\ref{eq_inner_elastic_boundary}) we introduced the tensor of magnetoelastic charges as follows:
\begin{equation}\label{eq_charges_general}
    \hat
    Q^{\mathrm{m-e}}\equiv-\frac{\lambda_\mathrm{anis}}{\mu}\left[\mathbf{L}_\mathrm{sur}\otimes\mathbf{L}_\mathrm{sur}-\mathbf{L}_{\mathrm{in}}\otimes\mathbf{L}_{\mathrm{in}}\right].
\end{equation}
For a rectangular-shaped sample the charges at the opposite edges
coincide: $\hat Q^{\mathrm{m-e}}(\mathbf{n}_1)=\hat
Q^{\mathrm{m-e}}(\mathbf{n}_3)$, $\hat
Q^{\mathrm{m-e}}(\mathbf{n}_2)=\hat
Q^{\mathrm{m-e}}(\mathbf{n}_4)$. We can express all the components
of the $\hat Q^{\mathrm{m-e}}$ tensor in terms of two nontrivial
combinations, $Q^{\mathrm{m-e}}_1\equiv
Q^{\mathrm{m-e}}_{XX}-Q^{\mathrm{m-e}}_{YY}$ and
$Q^{\mathrm{m-e}}_2\equiv2Q^{\mathrm{m-e}}_{XY}$ (in $X$, $Y$
coordinates).

\begin{figure}[htbp]
\begin{centering}
\includegraphics[width=0.5\textwidth]{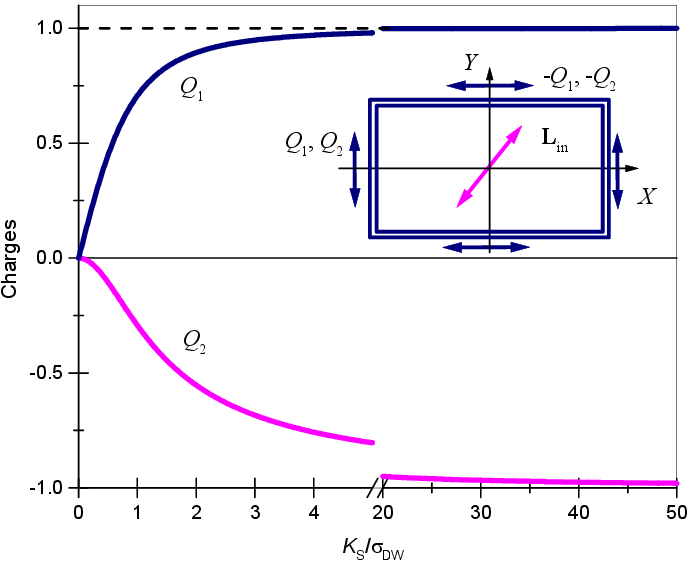}
  \caption{(Color online) Magnetoelastic charges $Q^{\mathrm{m-e}}$ (in $\lambda_\mathrm{anis}/\mu$ units) vs. surface constant $K_\mathrm{sur}$ calculated for single-domain state, $\psi=\pi/4$. Inset shows the charge distribution over the particles edges. Arrows indicate orientation of AFM vector at the surface and in the bulk.
 \label{fig_charges_vs_energy}}
 \end{centering}
\end{figure}

From definition (\ref{eq_charges_general}) and the relations
(\ref{eq_angle_boundary}) it follows that
\begin{equation}\label{eq_charges_specific_1}
 Q^{\mathrm{m-e}}_1(\mathbf{n}_{1,2})=\frac{\lambda_\mathrm{anis}}{\mu}\left(\cos2(\varphi_{\mathrm{in}}+\psi)-\frac{\sigma_{DW}\cos2(\varphi_{\mathrm{in}}+\psi)\mp K_\mathrm{sur}}{\sqrt{K_\mathrm{sur}^2\mp 2K_\mathrm{sur}\sigma_{DW}\cos 2(\varphi_{\mathrm{in}}+\psi)+\sigma_{DW}^2}}\right),
  \end{equation}
and
  \begin{equation}\label{eq_charges_specific_2}
 Q^{\mathrm{m-e}}_2(\mathbf{n}_{1,2})=\frac{\lambda_\mathrm{anis}}{\mu}\sin2(\varphi_{\mathrm{in}}+\psi)\left(1-\frac{\sigma_{DW}}{\sqrt{K_\mathrm{sur}^2\mp2K_\mathrm{sur}\sigma_{DW}\cos2(\varphi_{\mathrm{in}}+\psi)+\sigma_{DW}^2}}\right).
  \end{equation}

Magnetoelastic charges (\ref{eq_charges_general}) (as well as
(\ref{eq_charges_specific_1}), (\ref{eq_charges_specific_2}))
are similar to ``magnetostatic charges'' at the surface of FMs but
have another, magnetoelastic, nature (i.e., depend on
magnetostriction), and depend on the surface magnetic anisotropy
$K_\mathrm{sur}$. Magnetoelastic charges disappear in the limiting
case of small surface anisotropy $K_\mathrm{sur}\ll \sigma_{DW}$
and reach the maximum possible value when $K_\mathrm{sur} \gg
\sigma_{DW}$ (as illustrated in Fig.~\ref{fig_charges_vs_energy}).
Like magnetostatic, magnetoelastic charges depend on the
crystallographic orientation of the sample faces and vanish for
those parts of the surface where $\mathbf{L}_{\mathrm{in}} \|
\mathbf{L}_\mathrm{sur}$. From equations (\ref{eq_inner_elastic}),
(\ref{eq_inner_elastic_boundary}) it follows that magnetoelastic
charges produce \emph{long-range} (decaying as $1/r^2$) elastic
fields, which, in turn, lead to the ``destressing'' effects
(similar to ``demagnetising'' effects in FMs).

Another way to interpret the formation of
magnetoelastic charges presents itself in terms of incompatibility
of seed spontaneous deformations at the surface and in the
bulk. To this end, sufficient condition for chargesto  appear stems from the difference between the
surface and bulk values of \emph{any} physical quantity: magnetic (e.g.
nonmagnetic or paramagnetic surface), magnetoelastic, or
elastic (e.g. rigid shell).

\subsection{Average strains and shape-induced anisotropy}

\begin{figure}[htbp]
\begin{centering}
\includegraphics[width=0.5\textwidth]{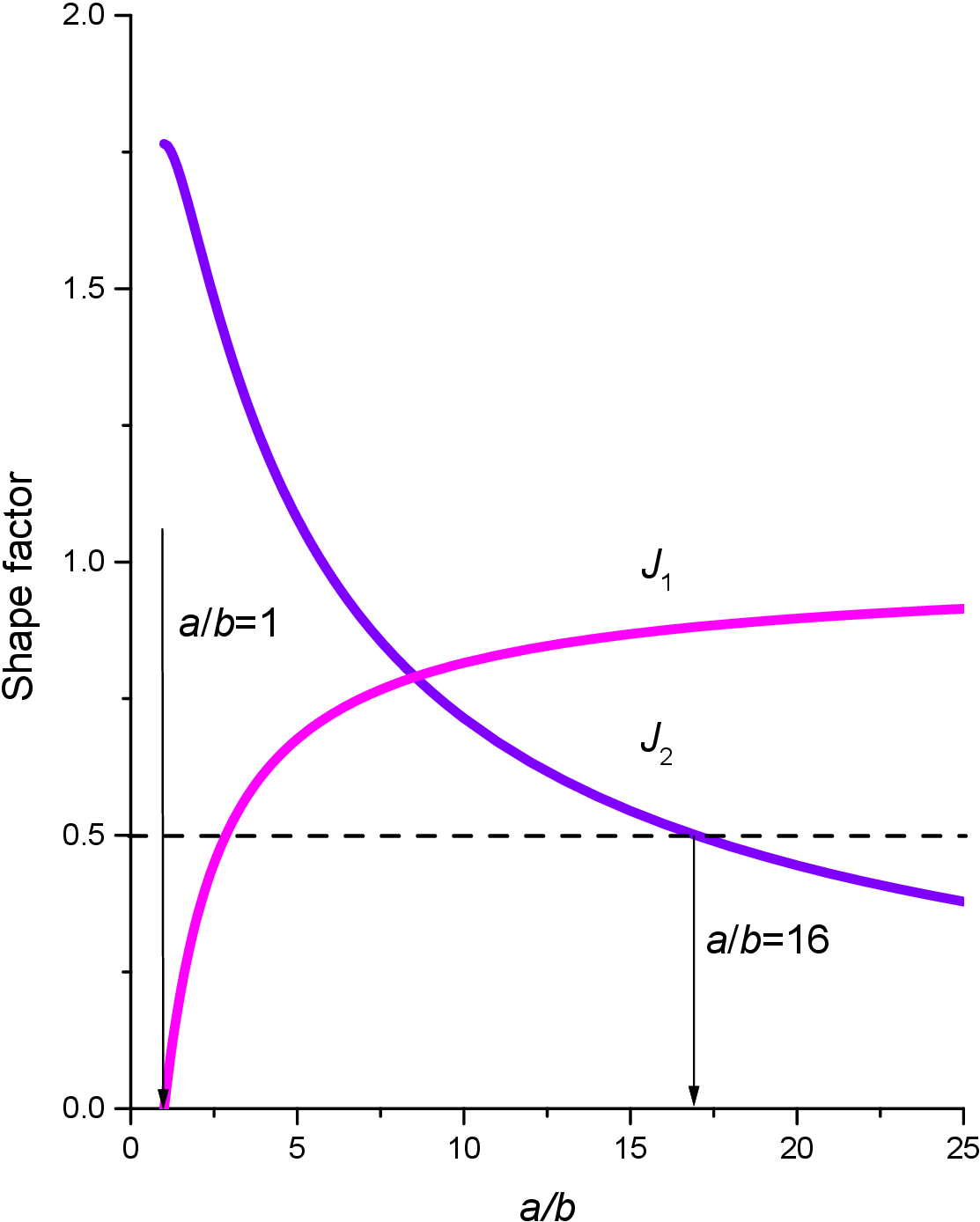}
  \caption{(Color online) Form-factors $J_1$, $J_2$ vs aspect ratio $a/b$. Arrows show the points where the functions
 $J_1$
($a/b=1$) and  $K_4^{\mathrm{sh}}$ ($a/b\approx 16$) change the
sign.\label{fig_shape_factor}}.
 \end{centering}
\end{figure}

At the next, \ref{item_three}, stage of the algorithm  we solve equations 
(\ref{eq_inner_elastic}), (\ref{eq_inner_elastic_boundary}) for
the displacement vector which, in general case, 
generates non-uniform field of additional (compared with
(\ref{eq_spontan-def})) elastic deformations. However, the main
contribution to the magnetic anisotropy comes from the shear
strains averaged over the sample volume (labeled as
$\langle\ldots\rangle$):
   \begin{eqnarray}\label{eq_average_deformation_2}
\langle
u_{XX}-u_{YY}\rangle&=&-\frac{\pi}{1+\nu_{\mathrm{eff}}}\left\{\left[
Q^{\mathrm{m-e}}_1(\mathbf{n}_2) +
Q^{\mathrm{m-e}}_1(\mathbf{n}_1)
\right]\left[1+\nu_\mathrm{eff}J_2\left(\frac{a}{b}\right)\right]\right.\nonumber\\
&-&\left.J_1\left(\frac{a}{b}\right)\left[
Q^{\mathrm{m-e}}_1(\mathbf{n}_1) -
Q^{\mathrm{m-e}}_1(\mathbf{n}_2) \right]\right\},
  \end{eqnarray}

   \begin{eqnarray}\label{eq_average_deformation_3}
2\langle u_{XY}\rangle&=&-\pi\left\{\left[
Q^{\mathrm{m-e}}_2(\mathbf{n}_2) +
Q^{\mathrm{m-e}}_2(\mathbf{n}_1)
\right]\left[1-\frac{\nu_\mathrm{eff}}{1+\nu_{\mathrm{eff}}}J_2\left(\frac{a}{b}\right)\right]\right.\nonumber\\
&-&\left.J_1\left(\frac{a}{b}\right)\left[
Q^{\mathrm{m-e}}_2(\mathbf{n}_1)- Q^{\mathrm{m-e}}_2(\mathbf{n}_2)
\right]\right\},
   \end{eqnarray}
where $J_1(a/b)$, $J_2(a/b)$ are the dimensionless shape functions
of the aspect ratio $a/b$ (see
Fig.~\ref{fig_shape_factor}):
\begin{equation}\label{eq_form_function_1}
J_1\left(\frac{a}{b}\right)=\frac{2}{\pi}\left[\arctan\frac{a}{b}-\arctan\frac{b}{a}+\frac{a}{4b}\ln\left(1+\frac{b^2}{a^2}\right)-\frac{b}{4a}\ln\left(1+\frac{a^2}{b^2}\right)\right],
  \end{equation}

\begin{equation}\label{eq_form_function_2}
J_2\left(\frac{a}{b}\right)=\frac{4}{\pi}\left[\frac{b}{a}\ln\left(1+\frac{a^2}{b^2}\right)+\frac{a}{b}\ln\left(1+\frac{b^2}{a^2}\right)\right].
  \end{equation}
Note that $J_2(a/b)=J_2(b/a)$; $J_1(a/b)=-J_1(b/a)$, so, $J_1=0$
for a square ($a=b$); in the opposite limiting case of high aspect
ratio ($a\gg b$) $J_1\rightarrow 1$,  $J_2\rightarrow 0$.

Substituting Exps.~(\ref{eq_charges_specific_1}),
(\ref{eq_charges_specific_2}), (\ref{eq_average_deformation_2}),
and (\ref{eq_average_deformation_3}) into
Eq.~(\ref{eq_system_mag}) we arrive at the following equation for
magnetic variable:
\begin{equation}
K_\perp\sin4\varphi_{\mathrm{in}}+K_2^{\mathrm{sh}}\sin2(\varphi_{\mathrm{in}}+\psi)+K_4^{\mathrm{sh}}\sin4(\varphi_{\mathrm{in}}+\psi)
= 0, \label{eq_system_mag_renorm}
\end{equation}
where we introduce the shape-dependent coefficients
$K_2^{\mathrm{sh}}$, $K_4^{\mathrm{sh}}$, and take into account
that $\Delta \varphi=0$.  In the limiting (and practically
important) case
$K_\mathrm{sur}\gg\sigma_{DW}$
\begin{equation}\label{eq_coef_shape}
K_2^{\mathrm{sh}}=2K^{\mathrm{m-e}}J_1\left(\frac{a}{b}\right),\qquad
K_4^{\mathrm{sh}}=K^{\mathrm{m-e}}\left[2J_2\left(\frac{a}{b}\right)-1\right],\qquad
K^{\mathrm{m-e}}\equiv\frac{4\pi\nu_{\mathrm{eff}}\lambda^2_\mathrm{anis}}{(1+\nu_{\mathrm{eff}})\mu}.
\end{equation}

In general case the coefficients $K_2^{\mathrm{sh}}$ and $K_4^{\mathrm{sh}}$ depend
on the constant $K_\mathrm{sur}$ of surface magnetic anisotropy
and vanish when $K_\mathrm{sur} \ll \sigma_{DW}$ (see
Appendix~\ref{sec_surf}).

Eq.~(\ref{eq_system_mag_renorm}) for the magnetic variables
$\varphi_{\mathrm{in}}$ can be treated as the minimum condition for
the effective energy density of the sample
\begin{equation}\label{eq_sample_energy}
  w_\mathrm{eff}\equiv\frac{W_\mathrm{eff}}{V}=-\frac{1}{4}\left[K_\perp
  \cos4\varphi_{\mathrm{in}}+\underline{2K_2^{\mathrm{sh}}\cos2(\varphi_{\mathrm{in}}+\psi)}+\underline{K_4^{\mathrm{sh}}\cos4(\varphi_{\mathrm{in}}+\psi)}\right],
\end{equation}
which, apart from the magnetic anisotropy, includes contributions
from magnetoelastic and surface forces (the underlined terms). The
last two terms in (\ref{eq_sample_energy}) cause the shape effects
in AFM nanoparticle. To illustrate this result we consider some
typical cases.

Let the sample edges be parallel to the easy magnetic axes
($\psi = 0$). In this case, as follows from
(\ref{eq_system_mag_renorm}) and (\ref{eq_sample_energy}), the
term with $K_2^{\mathrm{sh}}$ removes degeneracy of states
$\varphi_{\mathrm{in}} = 0$ and $\varphi_{\mathrm{in}} = \pi/2$.
This term is equivalent to uniaxial anisotropy, which selects the
state with the collinear orientation of AFM vectors at the surface
and in the bulk as energetically favorable. This means that the
AFM vector is parallel to the long edge of the rectangle: if $a>
b$, then $K_2^{\mathrm{sh}}> 0$  and $\mathbf{L} \| X$
($\varphi_{\mathrm{in}} = 0$). The second shape-induced term with
$K_4^{\mathrm{sh}}$ renormalizes the ``bare'' magnetic anisotropy
constant, $K_\perp \rightarrow K_\perp + K_4^{\mathrm{sh}}$;
however, this effect makes no influence on the orientation of the
AFM vector. For the square sample ($a=b$) the shape-induced correction
has the same sign as $K_\perp$ ($K_4^{\mathrm{sh}}> 0$) and thus
does not affect equilibrium orientation of the AFM vector. The
change of  $K_4^{\mathrm{sh}}$ sign appears for the 
samples with large aspect ratio ($a/b\approx 16$, see Fig.~\ref{fig_shape_factor}), where
uniaxial anisotropy governs the orientation of the AFM vector, and
shape-induced renormalization of $K_\perp$ is insignificant.

The role of the terms with $ K_4^{\mathrm{sh}}$ becomes noticeable
when the faces (edges) of the square ($a =b$) sample are cut at
the angle $\psi = \pi/4$ (i.e. along the ``hard'' magnetic axes). In this case
the uniaxial anisotropy vanishes, $K_2^{\mathrm{sh}} = 0$ and the
effective magnetic anisotropy constant decreases: $K_\perp
\rightarrow K_\perp-K_4^{\mathrm{sh}}$. Assuming that $ K_\perp $
and $ K_4^{\mathrm{sh}}$ have the same (spin-orbit) nature, we
conclude that the shape can change the direction of the easy axes
(if $ K_\perp <K_4^{\mathrm{sh}}$) or entirely compensate the 4-th
order magnetic anisotropy (if $ K_\perp \approx K_4^{\mathrm{sh}}
$), as it was recently observed in the experiments
\cite{Scholl:doi:10.1021/nl1025908}.

\section{Multidomain state, destressing energy and critical size}
In the multidomain state the seed distribution can, in principle, model the domains and domain boundaries both in the bulk and at the
surface. To simplify the problem we assume that distribution of the AFM vector
$\mathbf{L}_\mathrm{sur}(\mathbf{r})$ within each face is
homogeneous and $\mathbf{L}_\mathrm{sur}$ aligns due to the
surface anisotropy ($ K_\mathrm{sur} \gg \sigma_{DW} $), as shown
in Fig.~\ref{fig_near_surface} a. In this case, orientation of the AFM vector and, correspondingly, angle
$\varphi$, can
take one of two values within the bulk:
$\varphi_{\mathrm{in}}^I = 0$ or $\varphi_{\mathrm{in}}^{II} =
\pi/2$ (domains of two types, $I$ and $II$). At the surface
$\varphi_\mathrm{sur}^{(1)} = \varphi_\mathrm{sur}^{(3)}$ and
$\varphi_\mathrm{sur}^{(2)} = \varphi_\mathrm{sur}^{(4)}$.

Magnetoelastic charges appear near the surface (due to the
difference between $\mathbf{L}_\mathrm{sur}$ and
$\mathbf{L}_\mathrm{in}$) and at the domain walls in the bulk (due
to the difference between $\mathbf{L}_\mathrm{in}^I$ and
$\mathbf{L}_\mathrm{in}^{II}$). The total charge of the full
domain wall is zero because of the perfect compensation of the
charges with opposite signs. So, the field of internal charges
decreases rapidly with distance (as $1/r^6$, due to Coulomb-like
nature of the ``elastic'' forces) and can be neglected.

Near-surface domain structure generates two types of the charges,
$\hat Q^{\mathrm{m-e}}_{I}$ and $\hat Q^{\mathrm{me}}_{II}$,
corresponding to two types of the domains with
$\mathbf{L}_{\mathrm{in}}^I$ and $\mathbf{L}_{\mathrm{in}}^{II}$
(see Eq.~(\ref{eq_charges_general})). Thus, distribution of $\hat
Q^{\mathrm{m-e}}_{I,II}$ is space-dependent. We consider the
simplest case of the stripe domain structure (see discussion of possible
generalization below) and model it as
\begin{equation}\label{eq_charge_dependence}
    \hat Q^{\mathrm{m-e}}(\eta_j)=\langle \hat Q^{\mathrm{m-e}}_j\rangle +\left(\hat Q^{\mathrm{m-e}}_{I}-\hat Q^{\mathrm{m-e}}_{II}\right)f(\eta_j).
\end{equation}
Here $\eta_j$ is a local coordinate parallel to the $j$-th edge
of the sample (for example, $\eta_2=-X$ in
Fig.~\ref{fig_near_surface}), and $f(\eta_j)$ is a periodic
function with zero mean value:  $f(\eta_j+d)=f(\eta_j)$, $\langle
f(\eta_j)\rangle=0$; $d$ is a domain structure period.

In the case of the fine domain structure, $d \ll b, a$, the
averaged value $\langle \hat Q^{\mathrm{me}}_j \rangle$ is
independent of $j$ and coincides with that averaged over the particle
volume. As in the single-domain state, the effective contribution
from the averaged charges to the magnetic energy density is similar
to (\ref{eq_sample_energy}):
\begin{equation}\label{eq_sample_energy_destress}
  w_\mathrm{destr}=-\frac{1}{4}\left\{2K_2^{\mathrm{sh}}\langle\cos2(\varphi_{\mathrm{in}}+\psi)\rangle+K_4^{\mathrm{sh}}\left[\langle \cos2(\varphi_{\mathrm{in}}+\psi)\rangle^2-\langle \sin2(\varphi_{\mathrm{in}}+\psi)\rangle^2\right]\right\}.
\end{equation}
The term with $K_2^{\mathrm{sh}}$ corresponds to the uniaxial
shape-induced anisotropy. The second term, with
$K_4^{\mathrm{sh}}$, depends nonlinearly on the domain fraction
and is analogous to the demagnetisation energy of FM. Previously
\cite{gomo:PhysRevB.75.174439} we named this contribution as
\emph{destressing} energy, since it determines the equilibrium
domain structure in the presence of the external fields (in the defectless samples).

We estimate the energy contribution of the second term in
(\ref{eq_charge_dependence}) using the analogy between the theory
of elasticity and electro- (magneto-)statics: the total field of the alternating charge distribution with zero average
decays exponentially into the sample at
distances $d$: $u_{j}\propto \exp (-|X\pm a/2|/d), \exp (- |Y\pm
b/2|/d)$. The corresponding contribution to the total energy
density can easily be obtained by analogy with the well-known
Kittel expressions for FMs (formulae (54), (63) in
\cite{Kittel:RevModPhys_21_541}):
\begin{equation}\label{eq_near_surface_energy}
    w_{\mathrm{near-sur}}=A\mu \left(\hat Q^{\mathrm{m-e}}_{I}-\hat
    Q^{\mathrm{m-e}}_{II}\right)^2\frac{Sd}{V}
\end{equation}
where $A$ is a factor of the order of unity, $S$ is the surface
area.

Comparison of (\ref{eq_near_surface_energy}) and (\ref{eq_sample_energy_destress}) shows that $w_{\mathrm{near-sur}}/w_{\mathrm{destr}}\propto d/\ell\ll 1$  (where $\ell$ is the characteristic size sample). However,  contribution $w_{\mathrm{near-sur}}$, though small, 
defines the details of the domain structure (period, number of domains, orientation and shape of DW). Also, as
in the case of FM, a period of the equilibrium domain structure is
determined by the competition between the energy
(\ref{eq_near_surface_energy}) (which increases with $d$ increase) and
the total DW energy density
$w_{\mathrm{bound}} = \sigma_{\mathrm{DW}} \ell S / (Vd)$ (which decreases with $d$ increase). An
optimal value $d_{\mathrm{opt}}$ (up to an
unessential numerical factor) is
\begin{equation}\label{eq_domain_period}
  d_{\mathrm{opt}}\approx\sqrt{\frac{\ell\sigma_{\mathrm{DW}}}{\mu\left(\hat Q^{\mathrm{m-e}}_{I}-\hat Q^{\mathrm{m-e}}_{II}\right)^2}}.
\end{equation}

The period $d_{\mathrm{opt}}$ of the domain structure defines the
critical NP size $ \ell_{\mathrm{cr}}$, below which the formationnof AFM domains
 becomes energetically unfavourable:
\begin{equation}\label{eq_crit_size}
  \ell_{\mathrm{cr}}=d_{\mathrm{opt}}=\frac{\sigma_{DW}}{\mu\left(\hat Q^{\mathrm{m-e}}_{I}-\hat
  Q^{\mathrm{m-e}}_{II}\right)^2}.
\end{equation}

Let us compare  expressions (\ref{eq_domain_period}),
(\ref{eq_crit_size}) with the similar expressions for the FM samples
for two limiting cases.

\emph{Strong surface anisotropy}, $K_\mathrm{sur}\gg\sigma_{DW}$.
In this case, $\left(\hat Q^{\mathrm{m-e}}_{I}-\hat
Q^{\mathrm{m-e}}_{II} \right)\propto \lambda_{\mathrm{anis}}/\mu$
and
\begin{equation}\label{eq_crit_size_large}
  \ell_{\mathrm{cr}}=d_{\mathrm{opt}}=\frac{\sigma_{DW}}{\lambda^2_{\mathrm{anis}}/\mu}\propto \frac{\sigma_{DW}}{K_\perp}\propto\xi_{DW}.
\end{equation}
Here we used the fact that the magnetic anisotropy in the AFM has
the same nature as the magnetoelastic energy, resulting in
$K_\perp\propto \lambda^2_{\mathrm{anis}}/\mu$.

\emph{Weak surface anisotropy}, $K_\mathrm{sur}\ll\sigma_{DW}$. In
this case $\left(\hat Q^{\mathrm{m-e}}_{I}-\hat
Q^{\mathrm{m-e}}_{II} \right)\propto  \lambda_{\mathrm{anis}}
K_\mathrm{sur}/ (\sigma_{DW} \mu)$, and so,
\begin{equation}\label{eq_crit_size_small}
  \ell_{\mathrm{cr}}=d_{\mathrm{opt}}=\frac{\sigma_{DW}}{\lambda^2_{\mathrm{anis}}/\mu}\left(\frac{\sigma_{DW}}{K_\mathrm{sur}}\right)^2\propto\xi_{DW}\left(\frac{\sigma_{DW}}{K_\mathrm{sur}}\right)^2\gg \xi_{DW}.
\end{equation}

Thus, in AFMs, as opposed to FMs, the domain size and  the
critical particle size depend on the properties of the surface (in
this particular case -- on the magnetic surface anisotropy). In
the presence of strong surface anisotropy the characteristic
size of the domain is of the same order of magnitude as the DW
width. 
A similar result is obtained in
the FMs, provided that the magnetic anisotropy is of the same
order as the shape anisotropy. In the limiting case of zero
surface magnetic anisotropy the critical particle size tends to
infinity, in agreement with expected absence of the shape effects
in the large AFM samples (thermodynamic limit).

We emphasize that, in contrast to FMs, the equilibrium structure
of AFMs is formed by the orientational domains only (the angle
between vectors $\mathbf{L}$ in neighboring domains $<180^\circ$).
The translational 180$^\circ$ domains in collinear AFMs (that have opposite
 $\mathbf{L}$ directions) are physically
indistinguishable and should be identified by the presence of the interfaces.
This problem is out of the scope of the paper.

\section{Discussion}

We developed a model that takes account of the magnetic 
surface anisotropy and magnetoelastic interactions and predicts
the additional shape-dependent magnetic anisotropy in AFM. The surface
anisotropy selects one of the easy magnetic axes as energetically favorable, while  
magnetoelastic long-range interactions transfer the influence of
the surface on the entire NP bulk. Formally, we can describe such
effects using the tensor of magnetoelastic charges (\ref{eq_charges_general}) localized at the NP surface.

Shape-induced magnetic anisotropy manifests itself in two ways: \emph{i})
as the uniaxial anisotropy, which splits energy of otherwise degenerated 
equilibrium orientations of AFM vector; \emph{ii}) as a
``demagnetising'' (destressing) factor, which promotes 
formation of a certain domain type.

The first effect occurs when the shape-imposed easy axis is
perpendicular to the ``proper'' easy magnetic axis of the crystal (e.g.,
induced by an external magnetic field)
\cite{Scholl:d10.1021/nl300361e}. The constants of intrinsic and
shape-induced magnetic anisotropy are of opposite signs; so, there
is a critical aspect ratio $a / b$, at which spin-flop transition of the AFM vector takes
place.

The second effect appears when the domain structure is reversibly 
changed \cite{Lavrov:2002} under the action of external
fields (magnetic or mechanical). In the flat rectangular NP with
$a \ne b$, the shape-induced anisotropy plays the same role as the
external field, resulting in unbalance between 
domains of different types.

The constant of the surface anisotropy also determines the
critical parameters limiting formation of the domain structure. For example, if the NP size is
comparable with the domain structure period $d$, formation of the domain walls and thus of domain structure
is unfavourable. On the other hand, for the elongated samples with $a\gg d$ but $\left | K_2^{\mathrm{sh}} \left ((a / b)_{\mathrm{cr}} \right)
\right | \ge K_\perp$ (see (\ref{eq_sample_energy})),
 there is
only one possible equilibrium orientation of the AFM vector and thus only
one type of domains. In this case, the orientation of the easy axis depends
not only on ratio $a / b$, but also on the angle $\psi$, which
determines the orientations of the sample edges.
So, the control of the AFM particles shape allows not only to
create single-domain samples, but also to drive the magnetic
ordering direction.

\begin{figure}[htbp]
\centering{
   \includegraphics[width=0.5\textwidth]{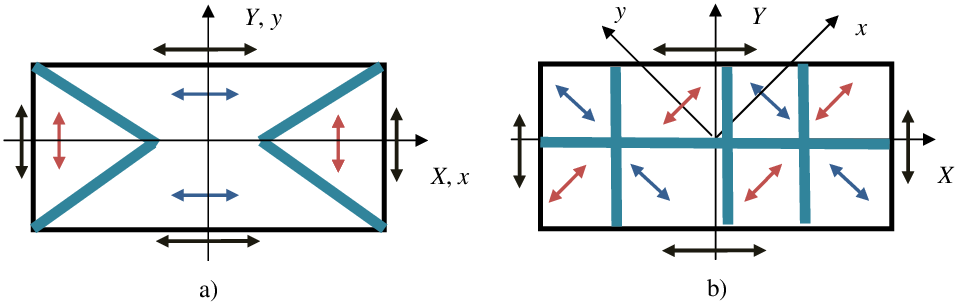}
\caption{(Color online) Multidomain state of the nanoparticle with
edges parallel (a) or at an angle 45$^\circ$ (b) relative to
crystal axes. Arrows outside the rectangle indicate the
orientation of the N\'eel vector at the surface.
 \label{fig_different_domain_structure}}
 }
\end{figure}

The magnetoelastic charges-based formalism allows to predict, at
least qualitatively, the morphology of the domain structure
depending on the size of the NP and crystallographic
orientations of its edges. Note that charge contribution 
increases the energy density of NP compared with the
case of an infinite crystal. So, charge-less (or zero mean) configuration is more favourable, as in
FMs. If the NP edges are parallel to the crystallographic axes,
the surface charges disappear in the structure that shown in
Fig.~\ref{fig_different_domain_structure} a, -- when the domain of
a certain type grows from the edge into the bulk as far as
possible. This type of the domain growth was observed
experimentally in \cite{Scholl:doi:10.1021/nl1025908}; the authors
named it ``edge effect''. Edge effect disappears if sample edges are rotated at angle $\psi
= \pi / 4$ with respect to easy magnetic axes.
Really, in this geometry charge vanishes only in average (due to 
formation of the domain structure that is periodic along the edge,
Fig.~\ref{fig_different_domain_structure} b). In this case, we
assume that domain formation starts from the vertices of the
rectangle, and the surface tension stresses can play a significant
role in this process. A detailed discussion of this issue is
beyond the scope of this paper.

The explicit form of the shape-induced anisotropy constants
(\ref{eq_coef_shape}), (\ref{eq_sample_energy}) depends on the
magnetic properties of the surface. In our model we suggest that
the magnetic ordering at the surface somehow differs from that
in the bulk, e.g. by orientation of easy axis (see
(\ref{eq_surface})). However, it is possible to generalize
the model and consider other typical situations: e.g., the surface
of the sample is paramagnetic, unlike the bulk. In this case, we
expect that the shape effects will show up in the destressing
energy (similar to (\ref{eq_sample_energy_destress})), but the
shape-induced anisotropy in the single-domain samples will be
absent, as well as edge effect. Thus, the shape effects give us
indirect information about the magnetic structure of the NP
surface.

The proposed model predicts the occurrence of the domains in
arbitrarily large samples, provided that $K_{\mathrm{sur}}\ne 0$.
Contribution of the magnetoelastic charges to the surface energy
is proportional to the sample volume and competes with the
anisotropy energy in samples of any size. On the other hand,
increasing the characteristic size of the sample, we can reduce
the influence of surface on the local properties up to the
thermodynamic limit. Thus, we can question the existence of the
upper critical size, above which the sample can be considered as
single-domain. To find a rigorous answer, we need to solve a
complicated problem, which is beyond the scope of our work; we confine
ourselfs to a few physical considerations.

Formally, we may move to the case of physically large samples (to
the thermodynamic limit) in two ways: either as ${\displaystyle
\lim_{K_{\mathrm{sur}}\rightarrow 0} \lim_{\ell \rightarrow
\infty}}$, or in the other order ${\displaystyle
\lim_{\ell\rightarrow\infty} \lim_{K_{\mathrm{sur}} \rightarrow
0}}$. In the first case, the surface leads to the shape effects
and the domain structure formation. Increasing the sample size,
and thus the domain size, we obtain large homogeneous regions, in
which the influence of domain walls and the surface can be
neglected (this issue is discussed in details below). In the second
case, we exclude the surface from consideration and get the
homogeneous throughout the sample solution (\ref{eq_spontan-def}),
which corresponds to the energy minimum. The domain structure is
absent and the size of the sample is not important as a
thermodynamic parameter.

We emphasize that our estimates of the domain structure period
(\ref{eq_domain_period}) and lower critical sample size
(\ref{eq_crit_size}) are based on the simplified Kittel's model of 
striped domain structure with one characteristic period.
While the optimal period is less than or equal to the critical
sample size (\ref{eq_crit_size}), such choice of seed 
distribution seems reasonable. However, if the sample
size (and $d_{\mathrm{opt}}$) increases, the contribution of the
charges $\hat Q_{I,II}^{\mathrm{me}}$ to the energy grows. At the
same time, energy can be decreased by the branching
(fractalisation) of the domain structure: the surface of ``large'' domain stimulates formation of small domains inside. Similar structures were observed in ferromagnetic and
ferroelastic materials (such as martensites, in which 
deformation is the primary order parameter \cite{Nishida200818}).
In \cite{Belokolos:0305-4470-34-11-324} authors show that the
scale invariance of the two-dimensional Laplace equation causes
the fractal nature of the ferromagnetic and intermediate state
superconducting structures. In our case, assuming the Coulomb
nature of the elastic forces, we can also expect that the system
of equations (\ref{eq_inner_elastic}) contains a similar (probably
more difficult) fractal solution. We suppose that a multi-domain hierarchical
structure, which contains ever smaller regions with various
orientations of the AFM vector, may also appear in large AFM samples. This leads us to the following
conclusions.

First, for large $\ell$ we need to adjust the estimate
(\ref{eq_domain_period}) for the domain structure period
$d_{\mathrm {opt}}$. Indeed, the total length of the domain walls
in the fractal structure increases with the domain size $d$
as $d^{D_H}$, where $D_H$ is the Hausdorf fractal dimension. Thus, the total energy of the domain walls changes as
$w_{\mathrm{bound}}\propto\ell d^{D_H-2}$ (similar estimate for
multiferroic BiFeO$_3$ was made in
\cite{Catalan:PhysRevLett.100.027602}), and the optimal domain
size is $d_{\mathrm{opt}}\propto\ell^{1/(3-D_H)}$. For the striped
domains $D_H=1$, which yields (\ref{eq_domain_period}); for
branching structures, obviously, $D_H>1$, and the dependence
$d_{\mathrm{opt}}(\ell)$ is stronger. Second, in the fractal
structure the ratio of the surface energy to the bulk energy
decreases with increasing $\ell$ slower than $1/\ell$, indicating
the important role of the surface in large samples.

Finally, we note that branching domain structure also allows transition to thermodynamic limit: as we have
already noted, for periodic structures the field of magnetoelastic
charges is screened over distances of the order of
$d_{\mathrm{opt}}$ from the surface. Thus, even in multidomain sample the local magnetic properties of homogeneous
regions (such as  orientation of AFM
vector, AFMR frequencies, susceptibility, etc.) depend
on the internal (bulk) parameters only, and the role of the surface energy is insignificant. 

Let us discuss another, practical, approach to the concept of the
upper critical dimension. Imagine that initially the multidomain
sample is transferred to a single-domain, homogeneous state
(without DWs) by an external field. The question is: will the
domain structure appear after the field is switched off? As in the
case of the FM materials, the answer depends on various parameters,
including the size of the sample, and the magnitude of the DW
formation activation barrier $U_{\mathrm{bar}}$. As we have
already noted, the domain formation starts at the surface -- from
edges or vertices of the particle, depending on the
crystallographic orientation of the surface. The domain nucleus
creates the elastic stress field; energy density of this field
decreases with distance (in analogy with the elastic energy of dislocation field) as $[(Q^{\mathrm{me}})^2/\mu]\ln r/r_0$ ($r_0$ is a
characteristic size of the order of the nucleus curvature radius).
If $U_{\mathrm{bar}} > [(Q^{\mathrm{me}})^2/\mu]$, then the domain walls preferably form in areas where the field of smagnetoelastic
charges  located at the opposite edges add contructively. Hence, we estimate the
upper critical size of the sample:
$\ell_{\mathrm{cr}}^{\mathrm{up}} \propto r_0
\exp{U_{\mathrm{bar}} \mu / (Q^{\mathrm{me}})^2}$. In small
particles, $\ell<\ell_{\mathrm{cr}}^{\mathrm{up}}$, the
interaction of charges located at the opposite edges is sufficient for
the DW formation. If $\ell >\ell_{\mathrm{cr}}^{\mathrm{up}}$, the
sample may remain in the metastable single-domain state.

Note that we have considered the ideal, i.e. defectless, sample,
eliminating the energy of twin boundaries and disclinations (the
latter inevitably arise in the areas of convergence of three or more domains),
and neglecting peculiarities of the AFM vector distribution near the vertices
of the rectangle. Certainly, these factors should influence the domain
structure formation and the effective magnetic anisotropy of the sample.
However, we guess that only the surface relates the internal magnetic properties
of NP and its form. We have shown that the shape effects can be caused by the 
long-range fields of non-magnetic nature -- elastic forces --  and so they should
appear in the ``pure'' AFM samples (without FM moment as well). 
The effects described above should be more pronounced in the small
(up to few critical lengths) samples: in this case, the formation
of the magnetic structure is determined mainly by the surface and
the influence of the defects can be neglected.

The results obtained show that the shape can be used  as a technological
factor which allows to drive, control and set the properties of
antiferromagnetic nano-sized samples.

The work is performed under the program of
fundamental Research Department of Physics and Astronomy, National
Academy of Sciences of Ukraine, and supported in part by a grant
of Ministry of Education and Science of Ukraine (N 2466-f).
\appendix
\section{Green tensor method for the displacement field calculation}\label{sec_Green}
Assuming that we know the distribution of the AFM vector
$\mathbf{L}(\mathbf{r})$, let us examine
Eqs.~(\ref{eq_system_el1})-(\ref{eq_system_el2}) for the elastic
subsystem. Corresponding boundary conditions at the surface are:
\begin{equation}\label{eq_surf_strain}
\frac{1+\nu_{\mathrm{eff}}}{1-3\nu_{\mathrm{eff}}}\mathbf{n}\mathrm{div}\mathbf{u}+\left(\begin{array}{c}
n_x(\nabla_xu_x-\nabla_yu_y)+n_y(\nabla_xu_y+\nabla_yu_x) \\
n_x(\nabla_xu_y+\nabla_yu_x)-n_y(\nabla_xu_x-\nabla_yu_y)
\end{array}\right)=-\frac{2\lambda_{\mathrm{anis}}}{\mu}\mathbf{L}_{\mathrm{sur}}(\mathbf{L}_{\mathrm{sur}}\mathbf{n}).
\end{equation}

To simplify, we skip the terms that describe the isotropic
magnetostriction (as insignificant for further discussion).

We denote the bulk forces vector by
\begin{equation}\label{eq_function_force}
   \mathbf{f}=-\left(\begin{array}{c}
   \nabla_x\cos2\varphi+\nabla_y\sin 2\varphi\\
   \nabla_x\sin2\varphi-\nabla_y\cos 2\varphi
   \end{array}\right),
\end{equation}
and the surface tension tensor (of the magnetostrictive nature) by
$2(\lambda_{\mathrm{anis}}/\mu)\mathbf{L}_{\mathrm{sur}}
\otimes\mathbf{L}_{\mathrm{sur}}$.

Let the functions $G_{kj}(\mathbf{r},\mathbf{r}^\prime)$
($k,j=x,y$) be the solutions of the equation:
\begin{equation}\label{eq_for_Green_1}
    \Delta G_{kj}(\mathbf{r},\mathbf{r}^\prime)+\nu_{\mathrm{eff}}\nabla_k \nabla_lG_{lj}(\mathbf{r},\mathbf{r}^\prime)=-\delta_{kj}\delta(\mathbf{r}-\mathbf{r}^\prime),
\end{equation}
with the following boundary conditions:
\begin{equation}\label{eq_for_Green_surf}
    (\mathbf{n},\nabla) G_{kj}(\mathbf{r}_{\mathrm{sur}},\mathbf{r}^\prime)+\nu_{\mathrm{eff}}\left[n_k \nabla_lG_{lj}(\mathbf{r}_{\mathrm{sur}},\mathbf{r}^\prime)+n_l \nabla_kG_{lj}(\mathbf{r}_{\mathrm{sur}},\mathbf{r}^\prime)\right]=0.
\end{equation}
Here, $\delta_{kj}$ is the Kronecker symbol, $\delta(\mathbf{r} -
\mathbf{r}^\prime)$ is the Dirac delta-function, $\mathbf{n}$ is
the surface normal in point $\mathbf{r}_{\mathrm{sur}}$.

The functions $G_{kj}(\mathbf{r},\mathbf{r}^\prime)$ coincide with
Green tensor for isotropic medium with fixed stresses (accurate
within constants). In this case, we can represent the displacement
vector as:
\begin{eqnarray}\label{eq_Green_Function_displ}
  u_j(\mathbf{r})=-\frac{2\lambda_{\mathrm{anis}}}{\mu}\int_V
  G_{kj}(\mathbf{r},\mathbf{r}^\prime)\nabla^\prime_l\left[L_k(\mathbf{r}^\prime)L_l(\mathbf{r}^\prime)\right]dV^\prime - \frac{2\lambda_{\mathrm{anis}}}{\mu}\oint_S
  G_{kj}(\mathbf{r},\mathbf{r}^\prime_{\mathrm{sur}})L_k(\mathbf{r}^\prime_{\mathrm{sur}})L_l(\mathbf{r}^\prime_{\mathrm{sur}})dS_l^\prime\nonumber\\
\end{eqnarray}

Substituing (\ref{eq_Green_Function_displ}) into energy expression
(\ref{eq_free_energy}) and taking into account boundary conditions
(\ref{eq_for_Green_surf}), we obtain elastic and magnetoelastic
energy contributions:
\begin{eqnarray}\label{eq_Green_Function_energy}
W_{\mathrm{add}}&=&\frac{2\lambda^2_{\mathrm{anis}}}{\mu}\int_V\int_V\nabla_m\left[L_j(\mathbf{r})L_m(\mathbf{r})\right]G_{kj}(\mathbf{r},\mathbf{r}^\prime)\nabla^\prime_l\left[L_k(\mathbf{r}^\prime)L_l(\mathbf{r}^\prime)\right]dVdV^\prime\nonumber\\
&+&\frac{2\lambda^2_{\mathrm{anis}}}{\mu}\oint_S\oint_S
L_j(\mathbf{r}_{\mathrm{sur}})L_m(\mathbf{r}_{\mathrm{sur}})G_{kj}(\mathbf{r}_{\mathrm{sur}},\mathbf{r}^\prime_{\mathrm{sur}})L_k(\mathbf{r}^\prime_{\mathrm{sur}})L_l(\mathbf{r}^\prime_{\mathrm{sur}})dS_mdS^\prime_l.
\end{eqnarray}

\section{Shape-induced contribution into the magnetic energy for an arbitrary constant $K_\mathrm{sur}$}\label{sec_surf}
In the general case, magnetoelastic charges
(\ref{eq_charges_specific_1}) and (\ref{eq_charges_specific_2})
depend on the ratio $s\equiv\sigma_{DW}/K_\mathrm{sur}$, which we
took as a unit when obtained Eqs. (\ref{eq_coef_shape}) and
(\ref{eq_sample_energy}). Here, we generalize these expressions
for arbitrary values of $s$.

Substituing (\ref{eq_charges_specific_1}),
(\ref{eq_charges_specific_2}) and
(\ref{eq_average_deformation_2}), (\ref{eq_average_deformation_3})
into equation (\ref{eq_system_mag}), we obtain expressions
(\ref{eq_system_mag_renorm}), where coefficients
$K_2^{\mathrm{sh}}$, $K_4^{\mathrm{sh}}$ depend on variables
$\varphi^{\mathrm{(in)}}$:
\begin{eqnarray}\label{eq_coef_shape_22}
K_2^{\mathrm{sh}}&=&K^{\mathrm{m-e}}\left[J_1\left(\frac{a}{b}\right)\frac{\Lambda_++\Lambda_-}{\Lambda_+\Lambda_-}
-\left(1+\nu_{\mathrm{eff}}J_2\left(\frac{a}{b}\right)\right)\frac{\Lambda_+-\Lambda_-}{\Lambda_+\Lambda_-}\right],
\end{eqnarray}

\begin{eqnarray}\label{eq_coef_shape_23}
K_4^{\mathrm{sh}}&=&K^{\mathrm{m-e}}\left[\left(2J_2\left(\frac{a}{b}\right)-1\right)\left(1-\frac{s(\Lambda_++\Lambda_-)}{2\Lambda_+\Lambda_-}\right)
-J_1\left(\frac{a}{b}\right)\frac{s(\Lambda_+-\Lambda_-)}{2\Lambda_+\Lambda_-}\right].
\end{eqnarray}
Here,
\begin{equation}\label{eq_designation}
 \Lambda_{\pm}\equiv\sqrt{1\pm2s\cos2(\varphi_{\mathrm{in}}+\psi)+s^2}.
\end{equation}

In the limiting case of the small magnetic anisotropy ($s\gg 1$)
both shape-dependent constants vanish:
\begin{equation}
K_2^{\mathrm{sh}}=K^{\mathrm{m-e}}J_1\left(\frac{a}{b}\right)\frac{2}{s}\rightarrow
0,\quad
K_4^{\mathrm{sh}}=-K^{\mathrm{m-e}}J_1\left(\frac{a}{b}\right)\frac{\cos2(\varphi_{\mathrm{in}}+\psi)}{s}\rightarrow
0.
\end{equation}

Equation (\ref{eq_system_mag_renorm}) may perform as the minimum
condition for the effective energy:
\begin{eqnarray}\label{eq_sample_energy_2}
  w_\mathrm{eff}&=&-\frac{1}{4}K_\perp
  \cos4\varphi_{\mathrm{in}}-\frac{1}{2 s}K^{\mathrm{m-e}}\left[J_1\left(\frac{a}{b}\right)(\Lambda_+-\Lambda_-)
+\left(1+\nu_{\mathrm{eff}}J_2\left(\frac{a}{b}\right)\right)(\Lambda_++\Lambda_-)\right]\\
  &-&\frac{1}{12s}K^{\mathrm{m-e}}\left[\left(2J_2\left(\frac{a}{b}\right)-1\right)\left(3s\cos4(\varphi_{\mathrm{in}}+\psi)+2(\Lambda_-+\Lambda_+)^3\right)
-2J_1\left(\frac{a}{b}\right)(\Lambda_--\Lambda_+)^3\right].\nonumber
\end{eqnarray}

%
\end{document}